\documentclass[twocolumn,trackchanges]{aastex62} %linenumbers

\graphicspath{{./}{figures/}}

\received{???}
\revised{???}
\accepted{???}
\shorttitle{Warp with LAMOST DR5 and Gaia DR3}
\shortauthors{Li \& Wang et al.}
\usepackage{amsmath}

\begin{document}

\title{Evidence for Populations-dependent vertical motions and the Long-lived Non-Steady Lopsided Milky Way Warp}
\author{Xiang Li}
\affil{Department of Astronomy, China West Normal University, Nanchong, 637002, P.\,R.\,China}
\author[0000-0001-8459-1036]{Hai-Feng Wang}
\affil{Department of Astronomy, China West Normal University, Nanchong, 637002, P.\,R.\,China}
\affil{CREF, Centro Ricerche Enrico Fermi, Via Panisperna 89A, I-00184, Roma, Italy}
\author{Yang-Ping Luo}
\affil{Department of Astronomy, China West Normal University, Nanchong, 637002, P.\,R.\,China}
\author{Mart\'in L\'opez-Corredoira}
\affil{Instituto de Astrof\'\i sica de Canarias, E-38205 La Laguna, Tenerife, Spain}
\affil{Departamento de Astrof\'\i sica, Universidad de La Laguna, E-38206 La Laguna, Tenerife, Spain}
\author{Yuan-Sen Ting}
\affil{Research School of Astronomy $\&$ Astrophysics, Australian National University, Cotter Rd., Weston, ACT 2611, Australia }
\affil{School of Computing, Australian National University, Acton ACT 2601, Australia}
\author{ \v{Z}ofia Chrob\'{a}kov\'{a}}
\affil{Faculty of Mathematics, Physics, and Informatics, Comenius University, Mlynsk\'a dolina, 842 48 Bratislava, Slovakia}

\correspondingauthor{HFW}
\email{hfwang@bao.ac.cn};\\

\begin{abstract}

We present the Galactic disk vertical velocity analysis using OB type stars (OB), Red Clump stars (RC), and Main-Sequence-Turn-Off stars (MSTO) with different average age populations crossed matched with LAMOST DR5 and Gaia DR3. We reveal the vertical velocities of the three populations varies clearly with the Galactocentric distance ($R$) and the younger stellar population has stronger increasing trend in general. The bending and breathing modes indicated by the vertical motions are dependent on the populations and they are varying with spatial locations. These vertical motions may be due to the Galactic warp, or minor mergers, or non--equilibrium of the disk. Assuming the warp is the dominant component, we find that the warp amplitude ($\gamma$, $Z_\omega$) for OB (younger population) is larger than that for RC (medium population) and the later one is also larger than that for MSTO (older population), which is in agreement with other independent analyses of stellar density distribution, and supports the warp is long-lived, non-steady structure and has time evolution. This conclusion is robust whether or not the line-of-nodes $\phi_w$ is fixed or as a free parameter (with $\phi_w$ is around 3$-$8.5$^{\circ}$ as best fit). Furthermore, we find that warp is lopsided with asymmetries along azimuthal angle ($\phi$).

\end{abstract}


\section{Introduction} 

Disk warp is a common asymmetrical structure in many disk galaxies \citep{Saavedra1990,Saavedra2003,Reshetnikov1998} and they have different shapes, including L-shape, S-shape and U-shape \citep{Kim2014}. \citet{Ann2006} observed 325 galaxies and found that 236 (73\%) of them showed warped structures, including 165 S-shaped (51\%) and 71 U-shaped (22\%), which might be caused by different formation mechanisms \citep{Saha2006}.

With the help of near-infrared sky survey, \citet{Guijarro2010} has observed 20 galaxies and found that 13 Galactic disks of which are warped. As one of the typical disk and spiral galaxies, the Milky Way also has a clear  warped disk which was firstly discovered by observations of neutral hydrogen (HI) \citep{Kerr1957,Westerhout1957,Bosma1981,Briggs1990,Nakanishi2003,Levine2006b}, it was then confirmed by observations of dust \citep{Freudenreich1994,Marshall2006}, the molecular clouds \citep{Grabelsky1987,Wouterloot1990,May1997} and different stellar tracers \citep{Momany2006,Reyle2009,Skowron2019a,Skowron2019b}. As seen from the view of the edge, the warp bends upward from the galactic disk plane to the north and downward on the other side \citep{Levine2006a} and the strength in the north might be greater than that in the south \citep{Skowron2019a}. To date the origin of the Milky Way warp is still mysterious and many works need to be done. 

Many mechanisms about the Galactic warp were proposed in recent years: the inflow of intergalactic matter into the halo (and consequent misalignment of halo-disk that produces the warp through gravitational interaction) \citep{Ostriker1989,Quinn1992,1999MNRAS.303L...7J,Bailin2003b} or directly onto the galactic disk \citep{Lopez2002a,2014MNRAS.440L..21H}, magnetic fields that exist between galaxies \citep{Battaner1998}, interaction between Sagittarius \citep{Bailin2003a} or Magellanic Clouds \citep{Burke1957,Weinberg2006} with the disk, the bending instability and self-excited warps or internally driven warps of the galactic disk \citep{Revaz2004,2022MNRAS.510.1375S}. 

The warp in various galactic disks might be a long-lived structure as mentioned in \citet{Rokar2010,Sellwood2013}. \citet{Lopez2014} analyzed the Milky Way warp based on the kinematic model and stellar tracers in the range of 5$-$16 \,kpc and suggested that S-shaped warp is a long-lived structure. However, \citet{Poggio2017} used OB stars to point out that the vertical motions of the Galactic disk cannot be explained by a stable long-lived warp model, and warp may be a transient structure or some phenomena acting on the gaseous component. There is not a strict limit for the long-lived and short-lived. But in this work we suggest the long-lived is more than few Gyr. For instance, less than 2 Gyr can be considered short-lived, whereas more than 5 Gyr can be considered long-lived and between 2 and 5 Gyr is intermediate.

\citet{wang2020a,wang2020b} have revealed the warp is a long-lived, nonsteady structure using the red clump stars with ages and clearly shown the younger stellar populations are stronger than the older ones. With the coming of high-precision data of Gaia EDR3, \citet{Chrob2022} used the star counts and the warp model $z_\omega = [C_{\omega}R({\rm pc})^{\epsilon_{\omega}}\sin(\phi - \phi_{\omega})]{\rm pc}$, where $C_{\omega}$ is the warp amplitude and $\phi_{\omega}$ is the Galactocentric angle defining the warp's line of nodes, to present that supergiants (younger populations) reach a maximum amplitude of $z_\omega$ = 0.658  kpc and a minimum amplitude of $z_\omega$ = $-$0.717 kpc at the distance $R$ = [19.5, 20] kpc, while the whole EDR3 populations (average old populations), reach a smaller maximum amplitude of $z_\omega$ = 0.360 kpc and a minimum of $z_\omega$ = $-$0.370 kpc. These results strongly support the points of \citet{wang2020b}, that is, the warp is a long-lived, but unstable structure or time-evolving. 

However, in contrast, \citet{Poggio2018} adopted two stellar tracers with different typical ages (upper main sequence and giant stars) to reveal that these two populations have similar kinematic characteristics. Similarly, \citet{ChengXinLun2020} explored the relations between stellar velocity and Galactocentric distance, angular momentum and azimuth, thus then unveiled that stellar vertical velocity with azimuth and prececession are basically similar in different age populations so they supported that the warp is from gravitational scenarios. These results are not consistent with \citet{Romero2019} results about warp kinematics using OB stars and RGB stars from Gaia DR2 to show the dependency of Galactic warp on age.

Recently, in order to deepen our understanding of the origin of warp, \citet{Chrob2021} also investigated more about the kinematics of warp based on the same data of Gaia DR2, with the warp model of \citet{Chrob2020} to re-calculate the warp precession ($\beta$). They discovered the value of warp precession $\beta$ = 4$^{+6}_{-4}$ km s$^{-1}$ kpc$^{-1}$, compatible with no precession at all. However, before that, \citet{Poggio2020} 's value of warp precession $\beta$ = 10.86$\pm$0.03 (stat.)$\pm$3.20 (syst.) km s$^{-1}$ kpc$^{-1}$ and the time-varying amplitude model can not fit the data well. Similarly, \citet{ChengXinLun2020}'s value of warp precession $\beta$ = 13.57$^{+0.20}_{-0.18}$ km s$^{-1}$ kpc$^{-1}$ with the data from Gaia DR2 and APOGEE \citep{Majewski2017}. 
According to \citet{Chrob2021}, the significant detections of precession are due to a wrong assumption of the amplitude of the warp independent of the stellar populations, which, when properly taken into account, it removes the necessity of precession.

As known, warp kinematics and dynamics as one of mechanisms to explain the disk oscillations, asymmetries and Galactoseismology, as such, the vertical disturbance on the galactic disk, which could be propagating in the form of bending waves. Based on two simulations in \citet{Khachaturyants2022}, one with warp caused by gas flowing into the galactic disk and the other without warp, they find that the simulation with warp can produce stronger bending wave. Better understanding of warp will definitely promote the understanding of the kinematic structures and gas dynamics of the galactic disk and Galactoseismology \citep{2021MNRAS.504.3168B,wang2018a,wang2018b,wang2019,wang2020a,wang2020b,wang2020c,Yu2021,wang2022c,Yang2022,2022arXiv220603495A}.

Assuming that the vertical motions could be contributed by warp, modelled as a set of circular rings that are rotated and whose orbit is in a plane with angle with respect to the Galactic plane, we have already made some progress of warp kinematics in \citet{Lopez2014,wang2020a,wang2020b}. In this work, different from our previous works, we only chose to use three different stellar populations (OB star, red clump star (RC) and main sequence turn off star (MSTO)) with distance of LAMOST DR5 (including DR4 MSTO) and Gaia DR3, to explore the warp again mainly from observational points of view in more details. 

This paper is structured as follows. Section 2 is about how we select samples and vertical velocity distributions in different stellar populations. Section 3 is about the model and method. We will show our results about the time-evolving warp in Section 4, and some discussion are in Section 5. Finally, we give the conclusions of this work.

\begin{figure}
  \centering
  \includegraphics[width=0.45\textwidth]{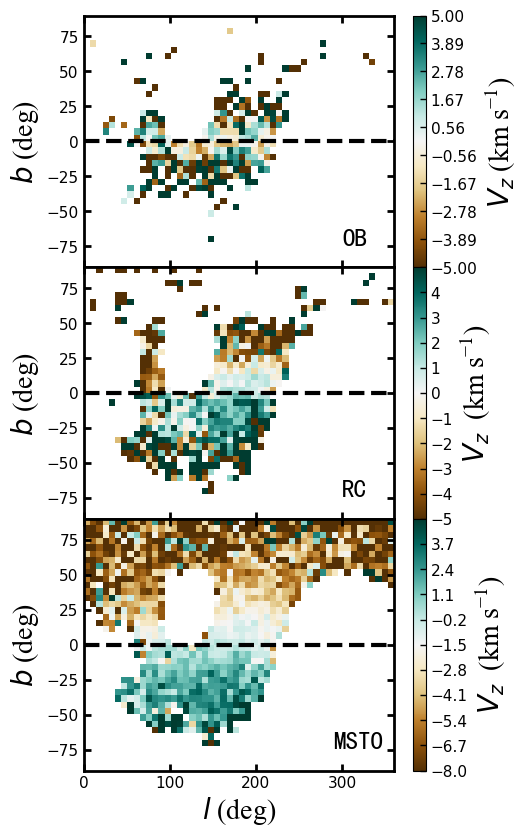}
  \caption{The vertical velocity of different tracers distributed in the celestial coordinates for longitude and latitude. From top to bottom, the panels are corresponding to OB, RC and MSTO samples.}
  \label{figure1}
\end{figure}

\begin{figure}
  \centering
  \includegraphics[width=0.45\textwidth]{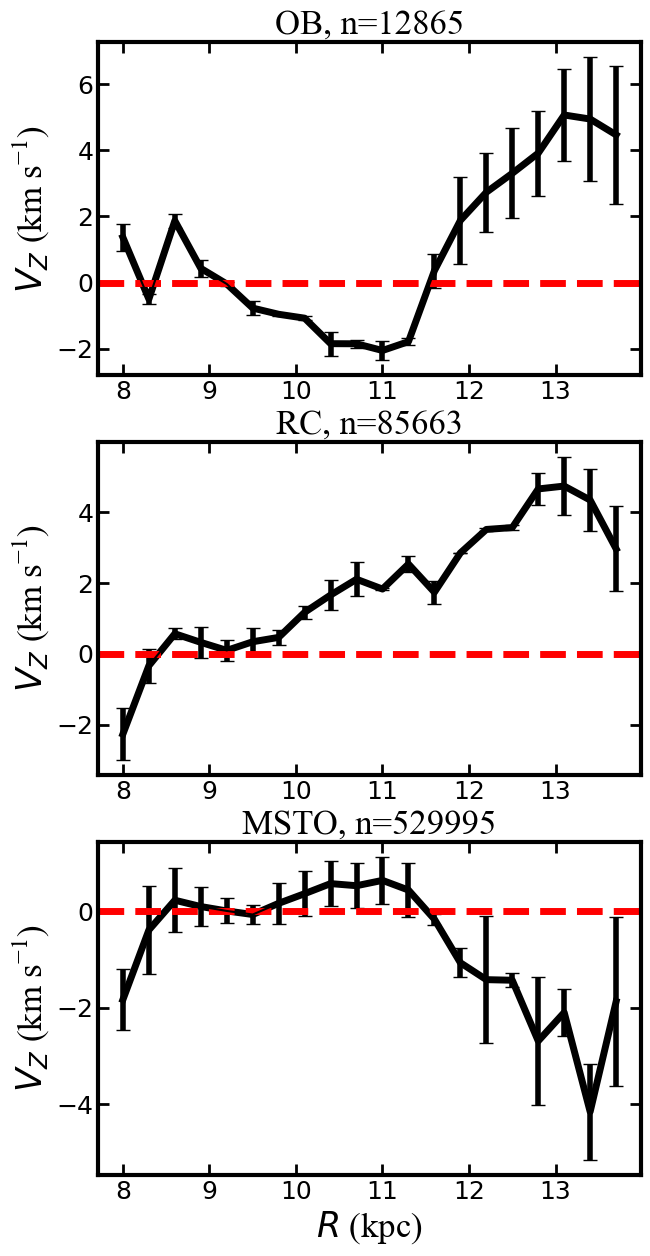}
  \caption{Vertical velocity distributions with radial distance for different stellar populations. The number of samples and the populations name are labeled at the top of each panel. The horizontal red dashed line is the zero of velocity value, used to guide our eyes. The three populations sample are from the LAMOST survey with different sampling rates and the azimuthal angle range of the three populations is mainly from $-$20 to 30$^{\circ}$.}
  \label{figure3}
\end{figure}

\begin{figure}
  \centering
  \includegraphics[width=0.45\textwidth]{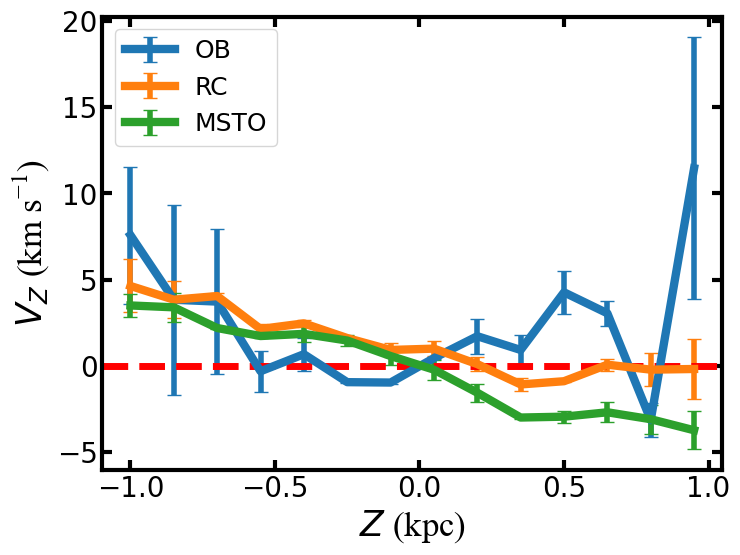}
  \caption{Vertical velocity distributions with vertical height for different stellar populations. Compared to Fig.~\ref{figure3} the warp signal is not clear at all but the vertical asymmetries in the north and south sides are shown here. So  difference in $V_{Z}$ is not driven by the different distribution in $Z$.}
  \label{Vzvsz}
\end{figure}

\begin{figure*}
  \centering
  \includegraphics[width=0.99\textwidth]{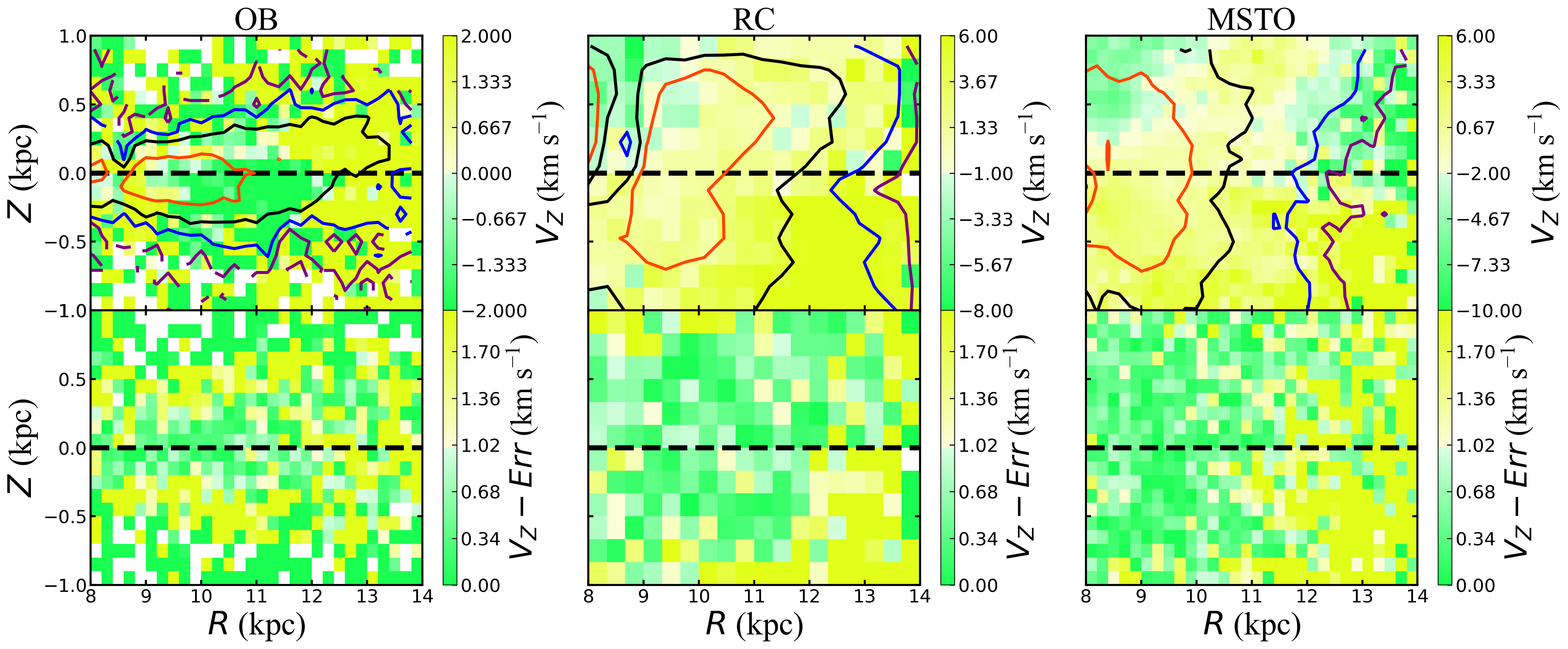}
  \caption{The upper three panels for the figure show the vertical velocity distributions of the disk in the $R$, $Z$ plane, the bottom three show the errors of the vertical velocity with bootstrapping. OB is in the left panel, the middle one is RC, and the MSTO is in the right panel. The abundant vertical motions or bending and breathing modes are showing here, the contours for star counts are also superimposed.}
  \label{figure2}
\end{figure*}

\section{Data} 

In this work, three different types of stellar tracers (OB, RC, MSTO) adopted are from the common stars of the LAMOST (The Large Sky Area Multi-Object Fiber Spectroscopic Telescope) Galactic spectroscopic surveys \citep{Cui2012,Deng2012,Zhao2012} and Gaia DR3 astrometric survey. The OB stellar samples can be obtained from \cite{Liu2019} and this catalog has been tested and widely used \citep{Cheng2019,wang2020c,Yu2021}, which has 14,344 OB stars. The distance in this catalog is from Gaia DR3 parallax  \citep{Gaia Collaboration2016,Gaia Collaboration2018}, since we only focus on the range of $R$ = 8$-$14 \,kpc the distance error is acceptable. RC sample details can be found in \citet{2018ApJ...858L...7T} and the uncertainty of distance is within 10\%, which has 175,202 RC stars. This catalogue is also widely used to explore the stellar mass and age \citep{2022arXiv220506144L}. MSTO samples and distances are based on \citet{Xiang2017a,Xiang2017b}, the error of distance is estimated to be 10\% to 30\%, which has 670,714 MSTO stars we choose and the ridge structure has been detailed studied with this sample \citep{wang2020c}.

Thanks to the latest Gaia DR3, we have obtained more accurate proper motions. Gaia DR3 catalog (both Gaia EDR3 and the full Gaia DR3) is based on data collected between 25 July 2014 and 28 May 2017 spanning a period of 34 months of data collection. As a comparison, Gaia DR2 was based on 22 months of data and Gaia DR1 was based on observations collected in the first 14 months of Gaia's routine operational phase. Gaia DR3 provides us with high-precision position, parallax, proper motion of 1.5 billion sources with a limiting magnitude of about G $\approx$ 21 and a bright limit of about G $\approx$ 3; and radial velocity of more than 33 millions sources with a limiting magnitude of G $\approx$ 14. The full astrometric solution has been done as 5-parameter solution for 585 million sources and as 6-parameter solution for 882 million sources. While the median uncertainties are 0.01$-$0.02 mas for G \textless \,15, 0.05 mas at G = 17, 0.4 mas at G = 20, and 1.0 mas at G = 21 mag. The uncertainty of proper motion is 0.02$-$0.03 mas yr$^{-1}$ for stars with G \textless \,15 mag, 0.07 mas yr$^{-1}$ for stars with G = 17 mag, 0.5 mas yr$^{-1}$ for stars with G = 20 mag, 1.4 mas yr$^{-1}$ for stars with G = 21 mag \citep{Gaia Collaboration2021,2022arXiv220800211G}.

By the following criteria we have the final sample for this work:

\begin{enumerate}
	\item {We eliminated the samples without parameters such as distance, radial velocity and proper motion.}
	\item {We only selected samples in the range of  $\left| Z \right|$  \textless \,1 kpc and  8 kpc \textless \,$R$ \textless 14 kpc.}
	\item {We choose samples with signal-to-noise ratio greater than 20, and [Fe/H] \textgreater \,$-$1.3 dex (For possible halo contamination).}
     \item {The three-dimensional velocity of the samples is in the following range, $V_R$ = [$-$150, 150] km s$^{-1}$, $V_\theta$ = [$-$50, 350] km s$^{-1}$, and $V_Z$ = [$-$150, 150] km s$^{-1}$.}
\end{enumerate}

The three-dimensional velocity we used is obtained by assuming the location of the Sun is $R_\odot$ = 8.34 kpc \citep{Reid2014} and $Z_\odot$ = 27 pc  \citep{Chen2001}. We use the \citet{Tian2015} solar motion values [$U_\odot$, $V_\odot$, $W_\odot$] = [9.58, 10.52, 7.01] km s$^{-1}$. The value of the circular speed of the LSR is 238 km s$^{-1}$ \citep{Schonrich2012}. Based on the Cartesian coordinate system, we calculate the 3D velocity, radial distance, vertical height and the azimuth $\phi$ of the stars using the method of $Galpy$ \citep{Bovy2015}. Notice that the different solar values and LSR will not change our conclusions. These kinematic parameters are also detailed described one by one in \citet{wang2018a,wang2019,wang2020a,wang2020b}. 

After above criteria, 529,995 MSTO stars, 85,663 RC stars and 12,865 OB stars are left and vertical velocity distributions in the celestial coordinate system (longitude and latitude), are shown in Fig.~\ref{figure1}. In this work, we only focus on stars with Galactocentric distance range from 8$-$14 \,kpc and vertical height $\left| Z \right|$  \textless \,1 \,kpc and in order to clearly present the variation of the vertical velocity ($V_Z$) of different stellar populations with the Galactocentric distance ($R$), we present the analysis displayed in Fig.~\ref{figure3} and the stellar population is denoted on the top of each panel. It can be seen from the figure that the vertical velocity of OB star increases from 2 km s$^{-1}$ to 6 km s$^{-1}$, the vertical velocity of RC increases from $-$2 km s$^{-1}$ to 5 km s$^{-1}$, and the vertical velocity of MSTO increases from $-$2 km s$^{-1}$ to 2 km s$^{-1}$ and then decreases by 6 km s$^{-1}$. The  kinematics distribution of the stellar populations is more or less showing the signal of warp and for the overall trend, the younger stellar populations increase faster than the older stellar populations, which is possibly implying younger stellar population has a stronger warp features than the older stellar populations. Similarly, vertical velocity distributions with vertical height for different stellar populations is shown in Fig.~\ref{Vzvsz}. Compared to Fig.~\ref{figure3} the warp signal is not clear but the vertical asymmetries in the north and south sides are shown here.

\section{Model} 

The vertical motions of (outer) Galactic disk may be related to the warp \citep{Rokar2010,Lopez2014,wang2018a,wang2020a,wang2020b,Lopez2020}.
As a first approximation, we will assume the observed vertical motions are mainly caused by the warp: a set of circular rings that are rotated, and whose orbit is in a plane with angle variable with time, $i_w(t)$, with respect to the Galactic plane. We can see more details in \citet{Lopez2014}. We will discuss later whether this approach fits the data or whether other elements [satellites interactions, non-equilibrium of the disk and so on \citep{wang2020a,Lopez2020}] are also necessary to explain the vertical velocites. Within this assumption, we can explore the amplitude evolution \citep{wang2020b} and precession \citep{ChengXinLun2020,Chrob2021} of warp by vertical velocity, or whether the vertical velocity field will be disturbed \citep{Drimmel2000}, and we will also examine the populations-dependent bulk motions.

For this work our model was established by \citet{wang2020b} which is similar to the model \citep{Lopez2014}:

\begin{align}\label{model}
V_Z(R \textgreater R_\odot) \approx& \frac{(R-R_\odot)^\alpha }{R}[\gamma \Omega_{\rm LSR}\cos(\phi - \phi_w) 
\notag
\\&+\dot{\gamma}R\sin(\phi - \phi_w)],
\end{align}
where $\phi_w$ is the azimuth of the line of nodes (deg) which is the azimuth with height of the warp equal to zero, $\gamma$ (kpc$^{-1}$) is the amplitude of the warp, and $\dot{\gamma}$ ($-$d($\gamma $)/d(age), kpc$^{-1}$ Gyr$^{-1}$) describes the warp amplitude evolution. We also assume a constant rotation speed $\Omega$($R$, $z$) = $\Omega_{\rm LSR}$ = 238 km s$^{-1}$ \citep{Schonrich2012}. This maybe slightly reduced for high $R$ or high $\left| z \right|$  \citet{Lopez2014}, but the order of magnitude does not change, and $V_Z$ is only weakly affected by a change of the rotation speed. $\alpha$ is a power law constant, $R$ and $\phi$ are from observational data points. 
 
Firstly, we assume $\alpha$ = 1 (no units) \citep{Reyle2009}, $\phi_w$ = $5^{\circ}$ (in the literature, the azimuth angle has a value between $- 28^{\circ}$ and  $18^{\circ}$ \citep{Chen2019, Lopez2002b, Momany2006, Reyle2009, Skowron2019b}). The results will be shown in the next part. Notice that the constant $\alpha$ = 2 (kpc$^{-1}$)  has also been tested in our study and the conclusions drawn from the analysis are stable, that is, the amplitude of the sample showed the same decreasing trend as shown in the next part. 

In this work, we use the sample with Galactocentric distance $R$ $\geq$ 8 kpc to obtain the best fitting results based on a  Markov  Chain  Monte  Carlo (MCMC) or EMCEE \citep{Foreman-Mackey2013}, we can obtain the likelihood distribution of the vertical velocity profile for the following fitting:

\begin{align}\label{lnlike}
&\mathcal{L} ({V_{\rm obs}(R_{i} |Z)} | \gamma,\dot{\gamma},\phi_w) = \prod\limits_{i}
\notag
\\&\exp\Big[ -\frac{1}{2} ((V_{\rm obs}(R_{i}|Z) - V_{\rm model}(R_{i}|Z,\gamma,\dot{\gamma},\phi_w))^2 \Big].
\end{align}                

When get the convergent parameters, the MCMC size is $50\times 3\times 10000$, and the step is 50. The values of each parameter($\gamma$, $\dot{\gamma}$ and $\phi_w$) based on MCMC fitting will be shown in the next section.

\section{Results} 

\subsection{Populations-dependent vertical bulk motions} 

The vertical velocity distributions of three samples in the $R$$-$$Z$ plane are shown in the Fig.~\ref{figure2}. It is clear that the vertical bulk motions are shown here, but different populations have different patterns. As seen for OB stars distribution, there is ``bending modes'' with negative vertical velocity between 10$-$12\,kpc and beyond 12\,kpc there is ``bending mode" with positive velocity, to our knowledge this is the first time to find this ``variable" vertical bulk motions in the young populations (Left panel). The positive vertical bulk motions or ``bending modes" is shown in the RC sample (middle panel) and in contrast, for MSTO population, we could see the positive ``breathing mode" between 8$-$9\,kpc and positive ``bending mode" between 9$-$12.5\,kpc and then the ``breathing mode" between 12.5$-$14\,kpc. In general,  ``bending mode" could be described as vertical velocity moves in the same direction and ``breathing mode" could be described as vertical motions move in the different direction, more details could be found in \citet{2014MNRAS.440.1971W}. The bending waves could be caused by warp with gas infall are shown in the  \citet{Khachaturyants2022}, which are being confirmed here by the bending and breathing modes from observations.

So we could claim here bending and breathing modes are dependent on the populations and they are evolving with spatial locations. The discovery of the populations-dependent vertical motions shown here is vital for us to better understand the Galactoseismology and we only address the warp contributions in this work.

\subsection{Warp Parameters $\gamma$, $\dot{\gamma}$ and $\phi_w$} 

\begin{figure*}
  \centering
  \includegraphics[width=0.9\textwidth]{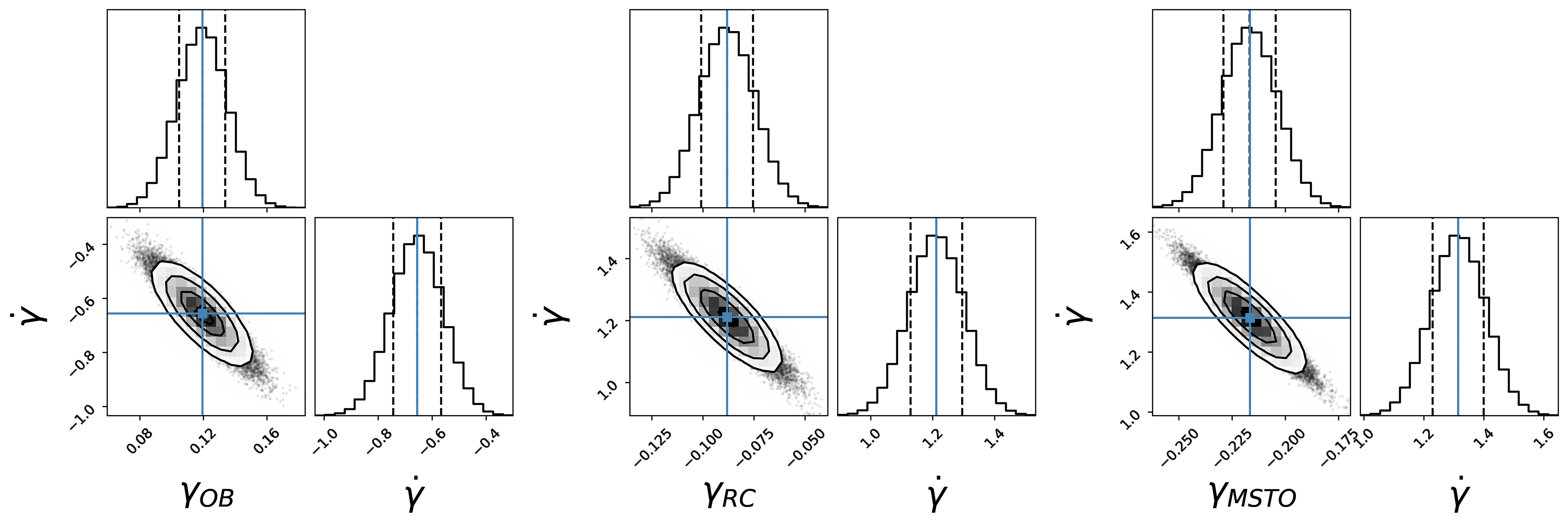}
  \caption{The likelihood distribution of the parameters ($\gamma$ and $\dot{\gamma}$) drawn from the MCMC simulation for each sample. From left to right, the panels are corresponding to OB, RC and MSTO. The warp amplitude $\gamma$ of younger populations is larger than that of older populations.}
  \label{figure4}
\end{figure*}

\begin{figure}
  \centering
  \includegraphics[width=0.45\textwidth]{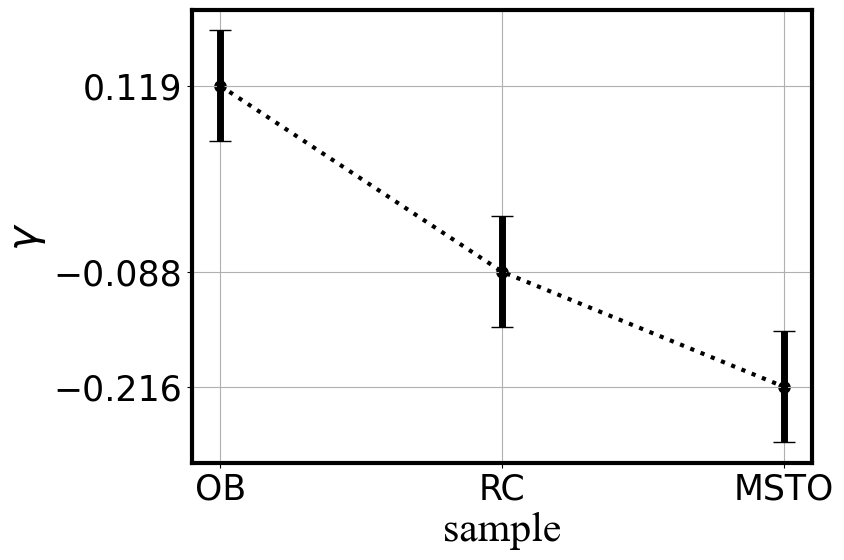}
  \caption{The warp amplitude ($\gamma$) of different populations obtained by MCMC fitting, the $x$-axis represents the sample name, and the vertical axis represents the warp amplitude. The error bar is given by bootstrap and the degeneracy of $\gamma$, $\dot{\gamma}$ are shown here. Negative $\gamma $ for some populations is not expected, implying that the line of nodes should be free parameter in fitting process.}
  \label{figure5}
\end{figure}

\begin{figure*}
  \centering
  \includegraphics[width=0.9\textwidth]{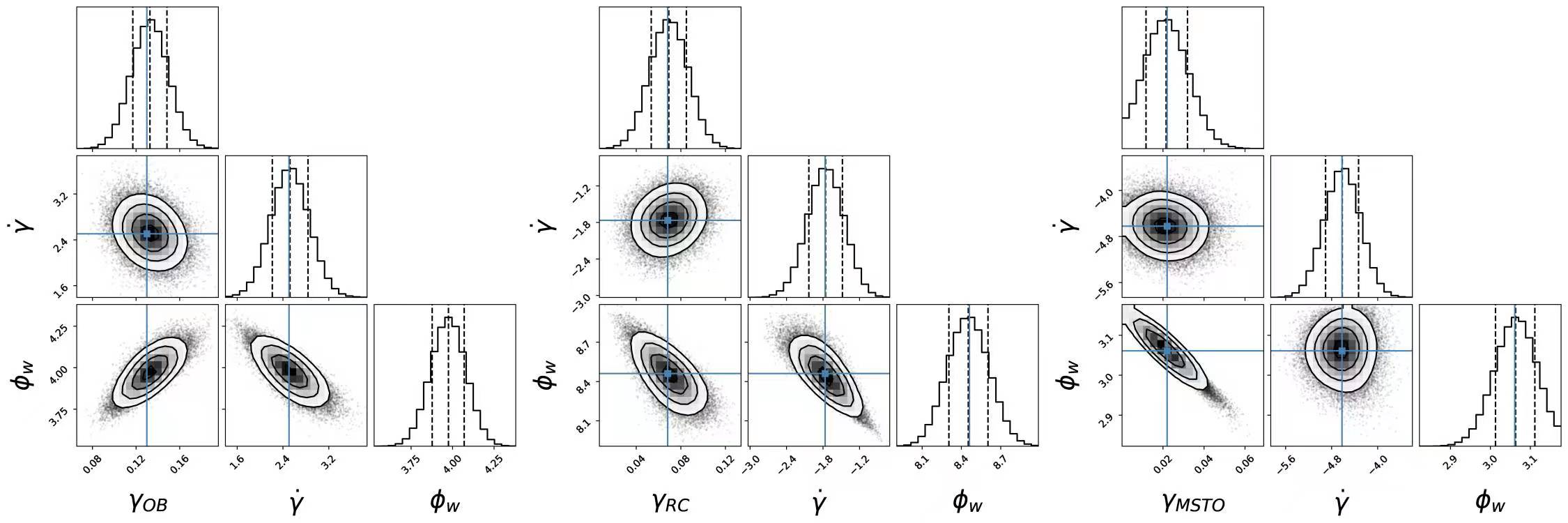}
  \caption{Similar to Fig.~\ref{figure4}, but the $\phi_w$ is free parameter, the likelihood distribution of the parameters ($\gamma$, $\dot{\gamma}$ and $\phi_w$) drawn from the MCMC simulation. The warp amplitude $\gamma$ of younger populations is larger than that of older populations. Here we could also see the degeneracy for $\gamma$, $\dot{\gamma}$ is broken and different than Fig.~\ref{figure4}. Note that $\dot{\gamma}$ is also different in different populations.}
  \label{figure6}
\end{figure*}

\begin{figure}
  \centering
  \includegraphics[width=0.45\textwidth]{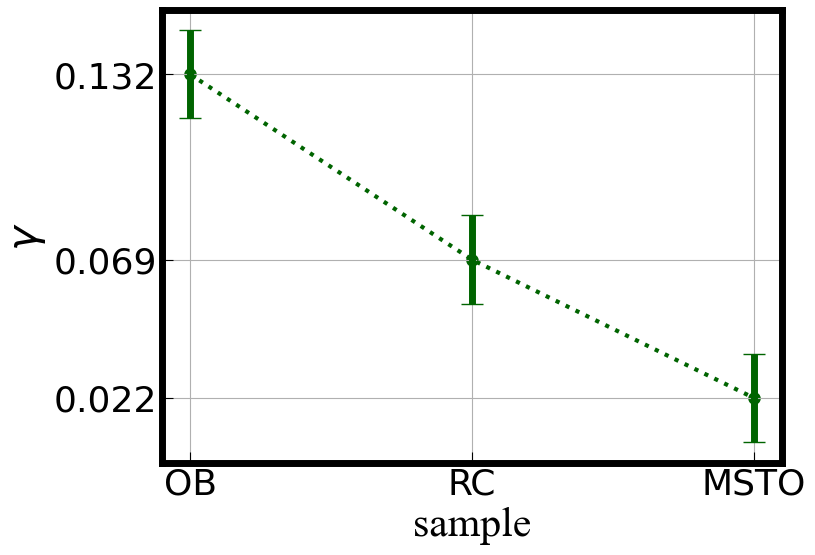}
  \caption{Similar to Fig.~\ref{figure5} with free parameter $\phi_w$, the decreasing pattern is similar to the Fig.~\ref{figure5}.}
  \label{figure7}
\end{figure}

\begin{figure}
  \centering
  \includegraphics[width=0.45\textwidth]{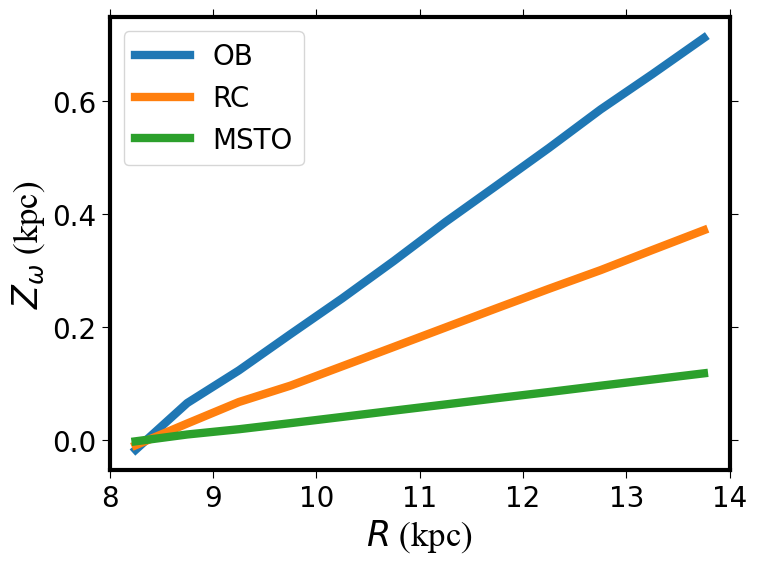}
  \caption{The maximum height caused by the warp is shown here, the $x$-axis is radial distance and $y$-axis could be seen as the warp amplitude.}
  \label{zw}
\end{figure}

\begin{figure*}
  \centering
  \includegraphics[width=0.9\textwidth]{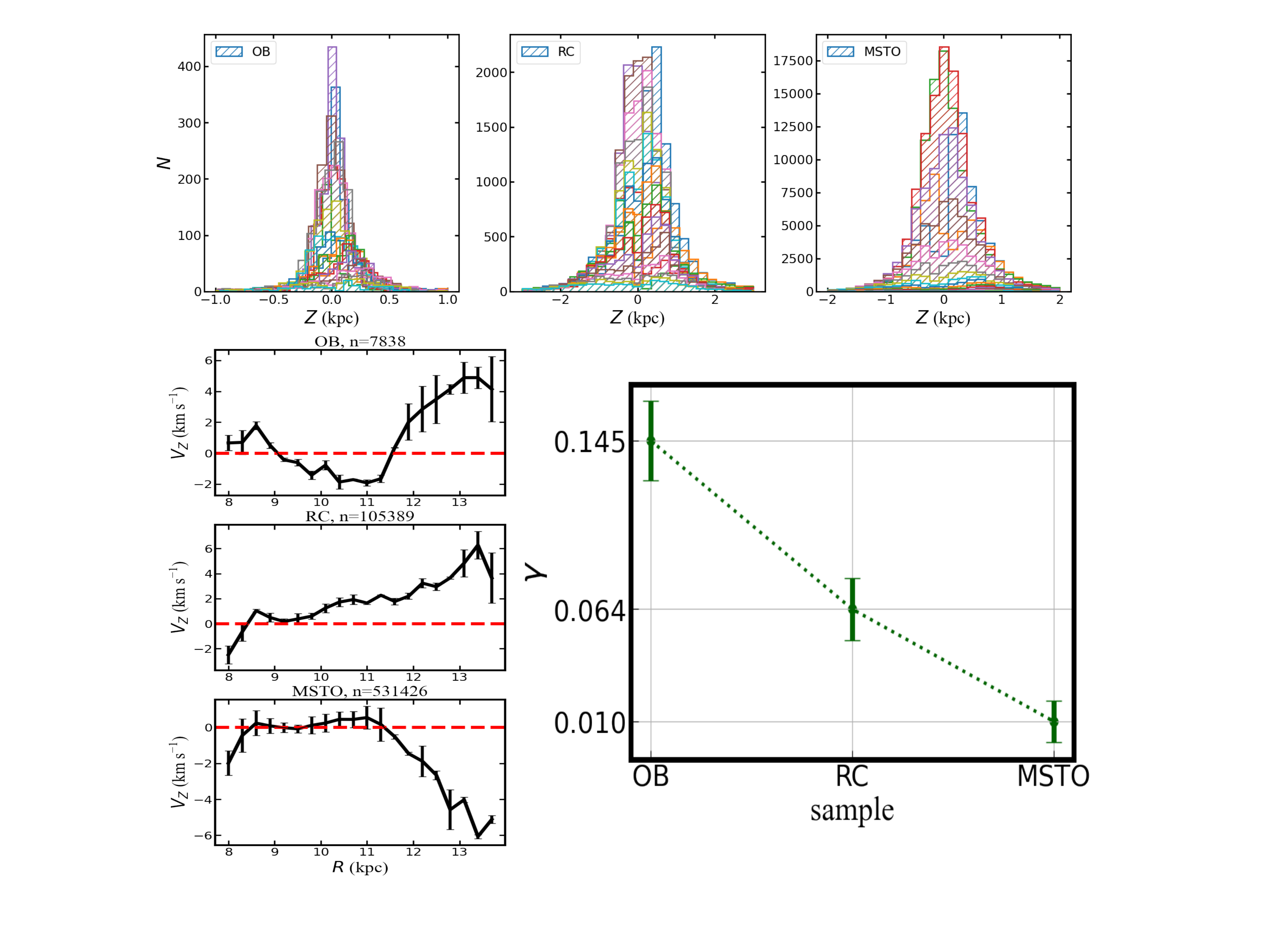}
  \caption{Top left: vertical heights distributions for different stellar populations in order to investigate the possible systematics mentioned in the Section 4.2. The peak values for height are: OB: $Z_0(R)$ = [0.0604, 0.0455, $-$0.022, 0.0170, 0.0133, $-$0.0055, 0.0052, 0.0074, 0.0159, 0.0181, 0.0255, 0.1117, 0.1442, 0.1642, 0.1837, 0.2156, 0.2186, 0.1426, 0.1523, 0.1120]; RC : $Z_0(R)$ =[0.4525, 0.4973, $-$0.2908, $-$0.1939,$-$0.0483, 0.0528, 0.1054, 0.1158, 0.1429, 0.1522, 0.1939, 0.1690, 0.2185, 0.2182, 0.2642, 0.2507, 0.1642, $-$0.1402, $-$0.2015, $-$0.2118]; MSTO: $Z_0(R)$ = [0.2709, 0.0714, 0.0072, 0.0356, 0.0437, 0.0491, 0.0553, 0.0725, 0.0979, 0.1783, 0.2589, 0.3433, 0.4197, 0.4944, 0.5243, 0.6012, 0.5973, 0.6271, 0.612, 0.4969]. Bottom left: vertical velocity distributions with radial distance for different stellar populations, similar to Fig.~\ref{figure3} but we select sample for $\mid$$Z$$-$$Z_0(R)$$\mid$ \textless 1 kpc for test, as shown, there are no large differences from the Fig.~\ref{figure3}. Similar to Fig.~\ref{figure7} but we select sample $\mid$$Z$$-$$Z_0(R)$$\mid$ \textless 1 kpc  for test, the decreasing pattern is the same and values are consistent within 1$\sigma$. The vertical mid-plane offsets are not the target of this work.}

 \label{appendixa.png}
\end{figure*}

Based on the vertical velocity of the sample we adopted, we conducted the first MCMC fitting under Eq. (\ref{model})  in the model. In this simulation, we assumed $\alpha$ = 1 \citep{Reyle2009} and the azimuth of the line of nodes $\phi_w$ = $5^{\circ}$, the fitting result is shown in Fig.~\ref{figure4} which is converged quite well. The amplitude of OB stars is $\gamma_{\rm OB}$ = 0.119 kpc$^{-1}$, amplitude of RC is $\gamma_{\rm RC}$ = $-$0.088 kpc$^{-1}$, amplitude of MSTO is $\gamma_{\rm MSTO}$ = $-$0.216 kpc$^{-1}$, which is also displayed in Fig.~\ref{figure5} with the error is given by bootstrap. Notice the negative $\gamma$ shown here is implying that pure sinusoidal warp cannot fit the data perfectly so that we need a more robust model in the future, but here we only focus on the relative amplitude difference, that is, younger populations are stronger than older ones. 

Furthermore, negative $\gamma $ for some populations is not expected when we fix the line of nodes, although we can not rule out the small possibility that it is meaning we detect the opposite warp sign which is expected in the south. Anyway, possibly our analyses indicate that the line of nodes is different for different populations, the clear negative degeneracy for $\gamma$ and $\dot{\gamma}$ will be broken when we set the line of nodes as free parameter, more analysis will be shown in Fig.~\ref{figure7}.

From Fig.~\ref{figure5}, we can see that the warp amplitude ($\gamma$) of different stellar populations decreases with the samples with different average ages. Note that, in this work, OB stars typical average age is around or less than few hundred\,Myr (\citet{wang2020c}) and RC is around 3.2\,Gyr and MSTO is definitely older than RC with 4.5\,Gyr. So the younger stellar population shows stronger warp amplitude than the older stellar population here. 

We naively assume $\phi_w$ = $5^{\circ}$ at first, but we know the azimuth of the line of nodes ($\phi_w$) obtained by different tracers varies from $-28 ^{\circ}$ to 18$ ^{\circ}$ \citep{Chen2019, Lopez2002b, Momany2006, Reyle2009, Skowron2019b}. Therefore, in order to explore whether or not $\phi_w$ will affect our conclusions in more details, we set the $\phi_w$ as a free parameter to fit again.

 As shown in Fig.~\ref{figure6}, we assume $\alpha$ = 1 \citep{Reyle2009}, and $\phi_w$ as a free parameter, MCMC fitting is carried out again. The amplitude values of warp are $\gamma_{\rm OB}$ = 0.132 kpc$^{-1}$, amplitude of RC is $\gamma_{\rm RC}$ = 0.069 kpc$^{-1}$, amplitude of MSTO is $\gamma_{\rm MSTO}$ = 0.022 kpc$^{-1}$, which is displayed in the Fig.~\ref{figure7}. Similarly, the $x$-axis represents the sample names and the $y$ axis represents the amplitude of different stellar populations acquired by model fitting. Obviously, the trend is consistent with Fig.~\ref{figure5}, as the increasing of age, the amplitude of warp is decreasing, the conclusion of which is also validated by the analysis in the Fig.~\ref{zw}, which is showing the maximum height calculated by the Eq. (3) in \citet{wang2020b}, but in this work we adopt the $\alpha$ = 1 not 2:
 
  \begin{equation}
Z_w(R > R_\odot,\phi=\phi _w+\pi /2)\approx \gamma (R-R_\odot )^{\alpha }
\label{zwmax}
,\end{equation}
 
The height is induced by the warp for different populations and displaying that, the younger the population is, the stronger the warp amplitude is. The values of $\dot{\gamma}$ for the three populations we obtained are not equal to 0, which proves that warp always exists but not stable. The azimuth of the line of nodes ($\phi_w$) of different stellar populations we fitted are about 3$-$8$^{\circ}$, and the change of $\phi_w$ is relatively small, which may be caused by the fact that the precession is too small to be detected in all warp models as mentioned in previous works \citep{Lopez2002a,Dubinski2009,Jeon2009}. Notice that the line of nodes could be expected to be straight within $R$ $\leq$ 4.5 disk scale lengths in some modelling works \citep{Shen2006, Joss2016}.

Since the selection is bounded by $\mid$$Z$$\mid$ \textless 1 kpc, one can wonder if this is introducing a possible systematic bias on the mean $V_Z$. So we select sample using $\mid$$Z$$-$$Z_0(R)$$\mid$ \textless 1 kpc, for which $Z_0(R)$ being the height of maximum density at radius $R$, to test this possibility as shown in Fig.~\ref{appendixa.png}. The results show the the influence is very minor, see more details in the caption.

\subsection{The Comparison of Model and Sample}

\begin{figure}
  \centering
 \includegraphics[width=0.45\textwidth]{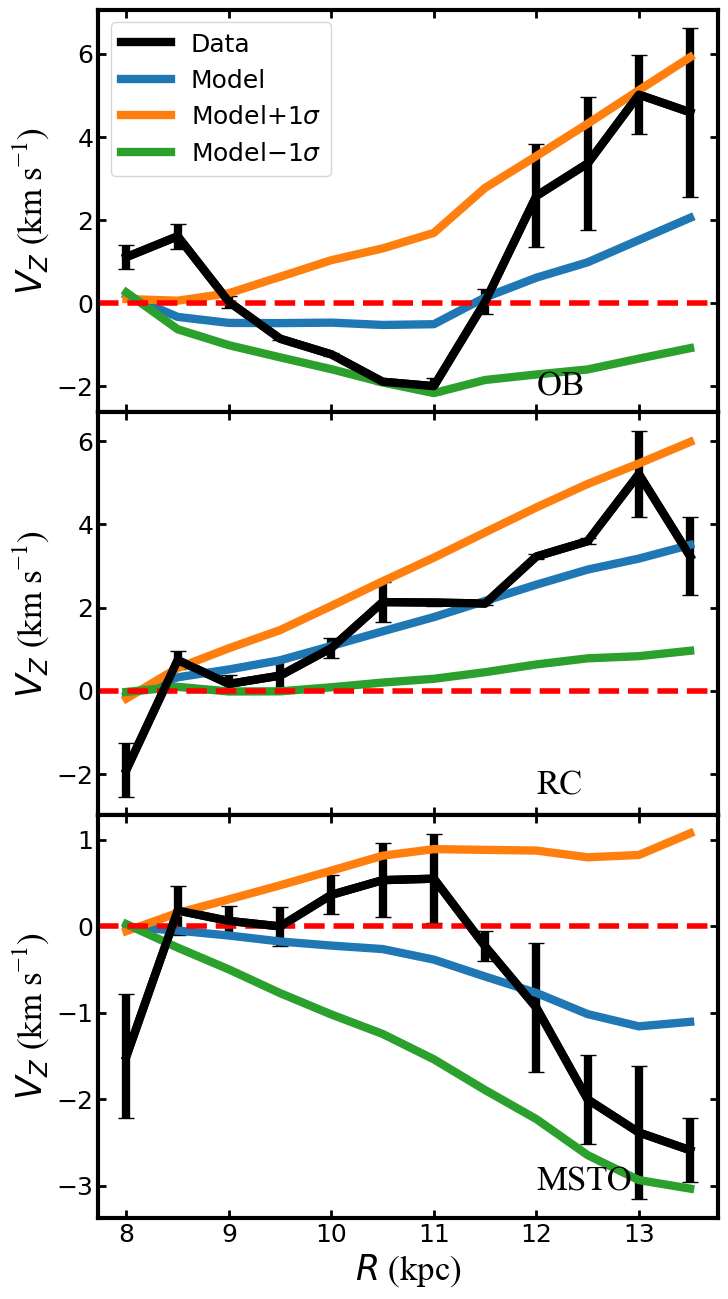}
  \caption{The one dimensional (1D) comparison for model and data, the model obtained by the MCMC fitting is the blue line and the orange is the model plus 1$\sigma$ and the green line is the model minus 1$\sigma$, black line is the data distribution. From top to bottom, corresponding to OB, RC and MSTO. The error bar is given by the bootstrap method and the model can be matched well within 1$\sigma$.}
  \label{figure8}
\end{figure}

In order to verify the results obtained by the model fitting, we compare the vertical velocity predicted by the model with the observed data. By using the warp parameters ($\gamma$, $\dot{\gamma}$ and $\phi_w$) obtained by model fitting we could have the mock data which could be compared with observations directly. As shown in Fig.~\ref{figure8}, the mean vertical velocity distributions as the function of the Galactocentric distance, the blue line is our model distribution, the yellow line is the model with $+$1$\sigma$, the green line is the model with $-$1$\sigma$, and the black line is our observational data, the error is given by bootstrap. From top to bottom panels, there are  three samples in this work respectively, namely OB, RC and MSTO stars. 

\begin{figure*}
  \centering
 \includegraphics[width=0.85\textwidth]{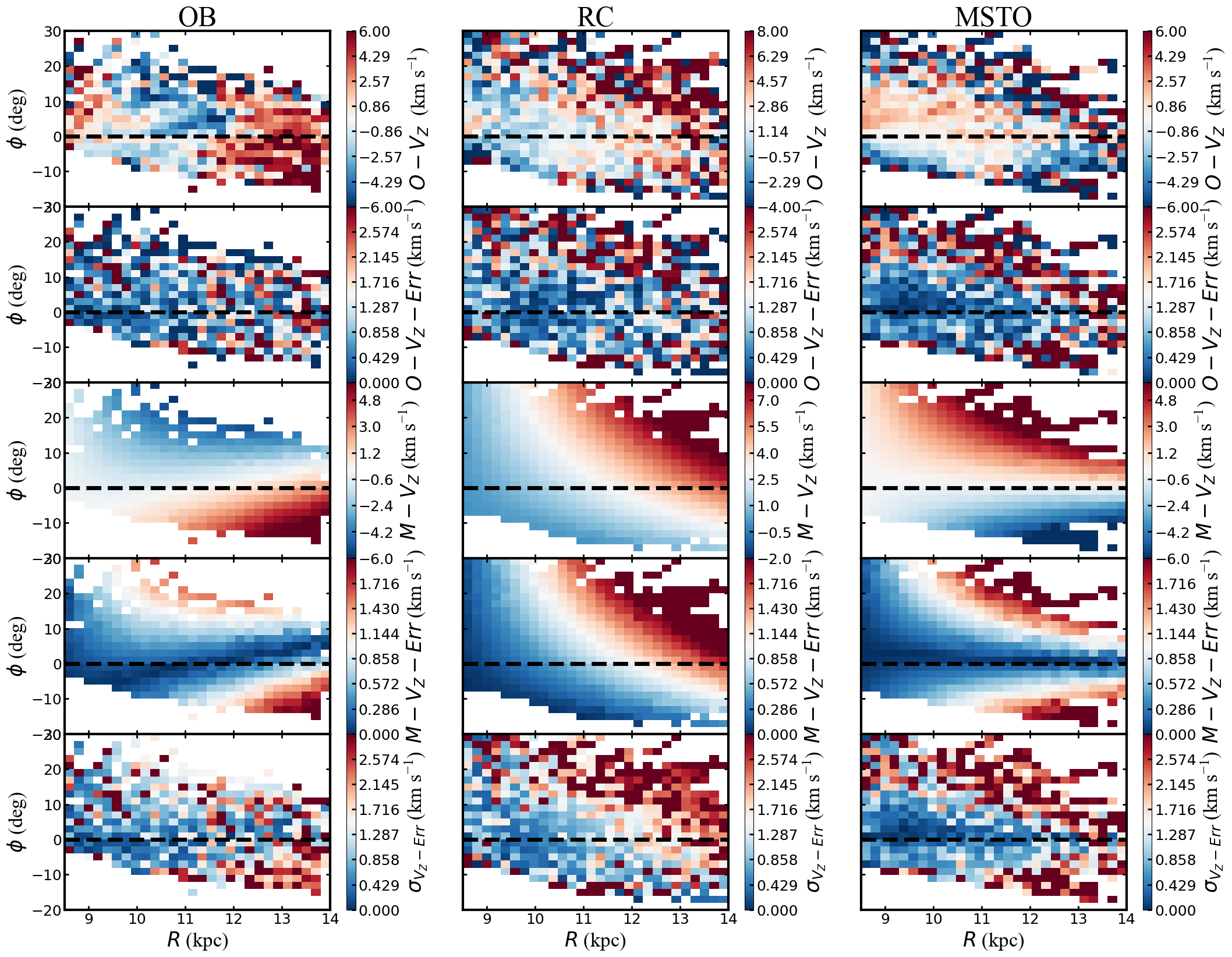}
  \caption{ The two dimensional (2D) comparison for model and data, the first row is the observational vertical velocity distribution on the $R$ and $\phi$ plane, the second row is the observational error with bootstrap, the third row is mock data with the model fitting, the forth row is error distributions of the model, the fifth row is the final error for our model and observations calculated by model error plus observational error with error propagation. From left to right columns, corresponding to OB, RC and MSTO. Notice that the quantitative match is not expected and qualitative comparison is acceptable here.}
  \label{RphiDataModel}
\end{figure*}

It can be seen from the Fig.~\ref{figure8} that our model can match well the overall trend of observed data at least in the 1$\sigma$ region. The two dimensional (2D) comparison for model and data is shown in the Fig.~\ref{RphiDataModel}, the first row is observational vertical velocity distribution on the $R$ and $\phi$ plane, the second row is observational error with bootstrap, the third row is mock data with the model fitting, the forth row is error of the model, the fifth row is the final error for our model and observations calculated by model plus observational error with error propagation. From left to right columns, corresponding to OB, RC and MSTO. Although we could see there are some difference between the model and data due to that the model is a simplified one and vertical motions might also be contributed by other mechanisms, the qualitative match is acceptable here.

\subsection{The Vertical Velocity as Functions of Azimuth}

\begin{figure}
  \centering
  \includegraphics[width=0.45\textwidth]{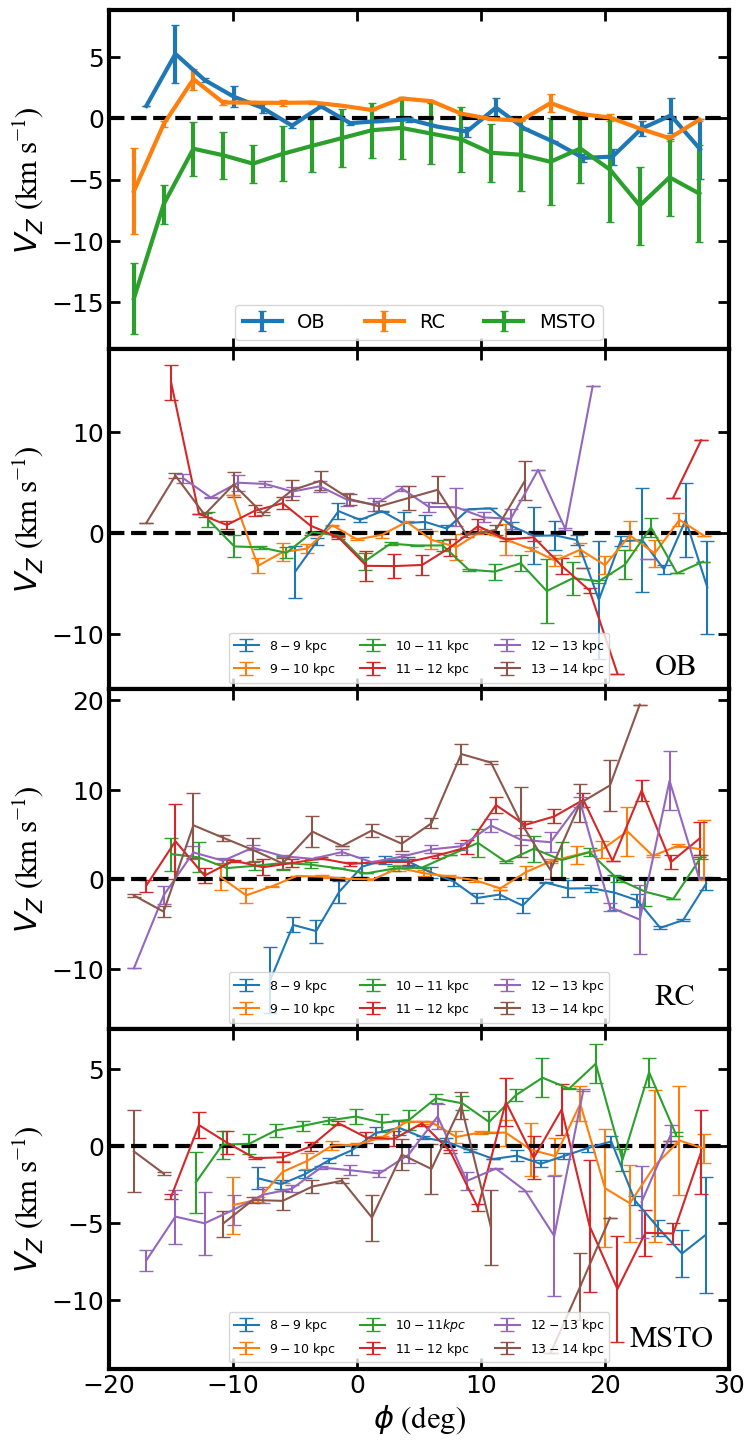}
  \caption{The vertical velocity distribution along with azimuth. The first panel shows the whole samples distributions. The bottom three shows the different samples distribution at different radial distance and from top to bottom are OB, RC and MSTO. The increase within $-10^{\circ}$ and the peak at $\phi \approx -10^{\circ}$, and then decrease with the azimuth as shown on the top panel, implying warp is lopsided.}
  \label{figure9}
\end{figure}

\citet{Romero2019} have used two different stellar populations from Gaia DR2 (OB stars and red giant branch stars) to find the asymmetry of the mean vertical distance between stars and Galactic plane, and suggest that warp is lopsided. Recently, \citet{ChengXinLun2020} combined Gaia DR2 astrometric solution, StarHorse distance and stellar abundances from the APOGEE survey to reveal the relations between the vertical velocity and azimuth. They found that the vertical velocity increases with the azimuth within 170$^{\circ}$ and reaches the peak at $\phi \approx 170^{\circ}$, and then decreases with the azimuth. The increasing and decreasing rates on both sides of the peak are obviously different, which is supporting the warp is lopsided.

Fig.~\ref{figure9} shows the vertical velocity of three stellar populations as the functions of azimuth. Top panel shows the distribution of vertical velocity of whole samples with azimuth. The blue line is OB, the yellow line is RC and the green line is MSTO, the error is given by bootstrap. As we could see, with the increase of azimuth, the vertical velocity also increases, reaches the peak at $\phi$ = $-$10$^{\circ}$ (in our definition for the anticenter direction,  $\phi$ = $-$10$^{\circ}$ is exactly the 170$^{\circ}$), and then decreases with the increase of azimuth, which are not symmetrical along the peak of vertical velocity.

In order to further explore the feature of warp, we divide the three samples into six groups with the distance ([8$-$9]\,kpc, [9$-$10]\,kpc, [10$-$11]\,kpc, [11$-$12]\,kpc, [12$-$13]\,kpc and [13$-$14]\,kpc), the distributions of the vertical velocity with the azimuth are shown in the bottom three panels of Fig.~\ref{figure9}, respectively. For the overall trend, we could see that the vertical velocity of the sample increases within $\phi$ = $-$10$^{\circ}$ and then, decreases with the increase of azimuth for MSTO, although the pattern is not as clear as the whole population. The asymmetry around $\phi$ = $-$10$^{\circ}$  is not clear for OB and RC but some other asymmetries are detected in different angle ranges, e.g.,  around $\phi$ = 10$^{\circ}$, all of which would point to some lopsidedness.

In short, based on the amplitude evolution, amplitude first derivative, vertical velocity with azimuth. For the assumption the vertical velocities are produced by the warp, we conclude that this warp is a long-lived, non-steady and lopsided structure; with an azimuth of the line of nodes around 3$-$8$^{\circ}$, consistent with other works. For the younger stellar population, it is greater than that of the older stellar population, which is consistent with the result of \citet{wang2020b} using red clump stars of different ages., which are possibly implying that the warp originated from gas infall onto the disk or other hypotheses that suppose that the warp mainly affects the gas, and consequently, younger populations tracing the gas are stronger than older ones.

\section{Discussion}

The warp contributes vertical nonaxisymmetric velocity patterns and wavelike density in the solar neighborhood and the outer Galactic disk of the Milky Way \citep{Widrow2012,Carlin2013,Williams2013,Xu2015,Pearl2017,Carrillo2018,wang2018a,wang2018b,Carrillo2019}, but to date we are still far from the exact knowledge of its origin.
The Milky Way disk coupling framework for population structure and Galactoseismology was mentioned in \citet{wang2018a,wang2020a}. Recently, using N-body simulations with different Sgr mass, \citet{2022ApJ...927..131B} clearly show that the Sgr is clearly not enough to cause the observed perturbation to the solar neighbourhood and multiple mechanisms are needed, which is consistent with our coupling framework. 

In recent years, many models such as the infall of gas, the influence of intergalactic magnetic field, and the interaction of nearby galaxies were proposed for the formation of warp (further details can be found in Section 1). According to \cite{Skowron2019b}, they attempt to divide them into two main scenarios: one is the non-gravitational mechanisms (the warp amplitude is dependent on the stellar age) dominated by, e.g., the gas infalling model (friction and collision), the other, e.g., is the gravitational mechanisms (the warp amplitude is not dependent on the stellar age) of the influence of satellite galaxies like the the Sagittarius Dwarf Spheroidal Galaxy (Sgr). 

In this work, three kinds of populations from spectroscopic surveys are used to represent different age populations, from younger to older with different average ages. Assuming the vertical velocities are produced by the warp, its signal would vary with the age shown in this work: younger one is stronger and warp has a clear evolution with age, which is strongly supporting the warp is not a transient one. In this work we attempt to conjecture that perhaps both mechanisms are contributing to the warp but there are different properties in the different populations and different phase-space, these coupling mechanisms framework is non-trivial to be scrutinized in the future. Notice also that to date we are still far away from the truth of the Sagittarius (Sgr) Dwarf Spheroidal Galaxy, we are not sure at all the gas contribution for it and its dance with the Milky Way \citep{wang2022b,wang2022a}. 

Some different observational results were reached by \citet{ChengXinLun2020}, who combined Gaia DR2 and Starhorse distance to explore the vertical and radial motion of stars in the Galactic disk and in the meantime, they established a warp model to support that warp is caused by the action of external gravity due to that there are no evolution for precession rate. However, \citet{Chen2019} used 1,339 classical Cepheids to explore the Galactic disk and found that the young stellar populations showed a stronger warp feature, which was also found by \citet{Chrob2020} with the help of Gaia DR2. More importantly, when we look carefully the pattern of vertical velocity vs. $\phi$ azimuth, we find our decreasing trend is weaker than \citet{ChengXinLun2020}, which will be confirmed by our Gaia DR3 new results, implying that the lopsided might be not so strong as \citet{ChengXinLun2020}. 

Nonetheless, vertical velocities alone cannot be used to derive the properties of the warp, because there are other possible dynamical causes for non-null vertical velocities \citep{Lopez2020}.
They might be generated by external perturbations or mergers, or by the fact the Galactic disk is a non-equilibrium system. The disk might have not reached equilibrium since its creation or because external forces, such as the Sagittarius dwarf galaxy, might perturb it, etc. We have found that a pure warp model with variable amplitude on time and epoch of formation of the stellar population fits reasonably well the observed features (see Fig. \ref{figure8}), which may indicate that the warp has an important average contribution to the vertical kinematics component. Moreover, details such as the exact dependence on the stellar population or the lopsidedness could be due to secondary factors as mergers, non-equilibrium or others. And one needs analyses different from the kinematics, mainly stellar density distribution (e.g., \citet{Chrob2022}) to corroborate that the warp is really dependent on the age. 

\section{Conclusions}

In short, our analysis about the warp has some similar results or implications (time-evolving warp for amplitude) to other previous works \citep{Liu2017b,Chen2019,Romero2019,wang2020a,wang2020b,Chrob2020}, but there are also differences from some others \citep{Poggio2018,Poggio2020, ChengXinLun2020}, which may be attributed to the different samples and models adopted in different works. Moreover, warp is lopsided, which has been mentioned in many works \citep{Amores2017,Poggio2018,Romero2019,ChengXinLun2020} and is also confirmed here, but due to that the azimuth range of our sample is smaller than $50^{\circ}$, we can not clearly show this feature as others. Furthermore, the asymmetry of the warp in the northern and southern disk is also an interesting point of view, which was presented in many works \citep{Momany2006,Reyle2009,Amores2017,Chrob2022}. We plan to continue to explore all of these in the future works using more samples and other warp models.

Based on the three different kinds of stellar populations (OB $\approx$ few to few hundred Myr, RC $\approx$ 3\,Gyr, MSTO $\approx$ 4\,Gyr) from the common stars of  LAMOST DR5 and Gaia DR3, we investigate the evolution of warp amplitude, the azimuth of the line of nodes, and the changes of vertical velocity with the azimuth angles. We clearly find the warp amplitude, obtained from the kinematic model, of the younger stellar populations is larger than that of the older stellar populations with temporal evolution. Moreover, the azimuth of the line of nodes is about 3$-$8$^{\circ}$ and the vertical velocity as a function of azimuth which is supporting that the warp is lopsided but perhaps not as strong as the results of recent APOGEE work. Nonetheless, these conclusions are conditioned to a predominance of the warp in the vertical velocities,
and that the effect of minor merger or non-equilibrium is much smaller.

Furthermore, we firstly discover that the vertical bulk motions are different in different populations, the bending and breathing modes are populations-dependent, which provide more constraints on the theoretical models and invite theorist of community to investigate more about the disk bending and breathing, and the relations to warp, mergers or non-equilibrium scenarios.

This analysis of vertical velocities in term of a warp, together with other independent analyses with different methods and
data, favour a scenario in which the warp is a long-lived non-stationary lopsided structure. As prospects, in the future work, we will make full use of different kinds of stellar samples with better sampling and sky coverage and more different models to unveil more mysteries of the S-shape Milky Way warp, more Gaia DR3 works will be finished in the future.

\acknowledgements
We would like to thank the anonymous referee and Francesco Sylos Labini for his/her very helpful and insightful comments. We acknowledge the National Key R \& D Program of China (Nos. 2021YFA1600401 and 2021YFA1600400). HFW acknowledges the support from the project “Complexity in self-gravitating systems” of the Enrico Fermi Research Center (Rome, Italy) and science research grants from the China Manned Space Project with NO. CMS-CSST-2021-B03, CMS-CSST-2021-A08. L.Y.P is supported by the National Natural Science Foundation of China (NSFC) under grant 12173028, the  Chinese Space Station Telescope project: CMS-CSST-2021-A10, the Sichuan Science and Technology Program (Grant No. 2020YFSY0034), the Sichuan Youth Science and Technology Innovation Research Team (Grant No. 21CXTD0038), Major Science and Technology Project of Oinghai Province (Grant No. 2019-ZJ-A10), and the Innovation Team Funds of China West Normal University (Grant No. KCXTD2022-6). Y.S.T. acknowledges financial support from the Australian Research Council through DECRA Fellowship DE220101520. H.F.W. is enthusiastic  for the plan ``Mapping the Milky Way (Disk) Population Structures and Galactoseismology (MWDPSG) with large sky surveys" in order to establish a theoretical framework in the future to unify/partly unify the global picture of the disk structures and origins with a possible comprehensive distribution function. The Guo Shou Jing Telescope (the Large Sky Area Multi-Object Firber Spectroscopic Telescope, LAMOST) is a National Major Scientific Project built by the Chinese Academy of Sciences. Funding for the project has been provided by the National Development and Reform Commission. LAMOST is operated and managed by National Astronomical Observatories, Chinese Academy of Sciences. This work has also made use of data from the European Space Agency (ESA) mission {\it Gaia} (\url{https://www.cosmos.esa.int/gaia}), processed by the {\it Gaia} Data Processing and Analysis Consortium (DPAC, \url{https://www.cosmos.esa.int/web/gaia/dpac/consortium}). Funding for the DPAC has been provided by national institutions, in particular the institutions participating in the {\it Gaia} Multilateral Agreement.

\end{document}